\documentclass{elsart}
\usepackage{graphicx}
\usepackage{dcolumn}
\usepackage{bm}
\usepackage{amssymb}
\usepackage{amsmath}

                              \newlength{\strikewidth}
                              \newlength{\strikelength}
\begin{document}
\begin{frontmatter}
\title{Solar System tests disfavor $f(R)$ gravities}

\author{Xing-hua Jin}
\author{Dao-jun Liu}
\author{Xin-zhou Li}\ead{kychz@shnu.edu.cn}
\address{Center for astrophysics, Shanghai Normal University,
100 Guilin Road, Shanghai, 200234, China\\
School of Science, East China University of Science and Technology,
130 Meilong Road, Shanghai, 200237, China}


\begin{abstract}
Using the elegant method employed recently by Erickcek, Smith and
Kamionkowski \cite{Erickcek}, on the premise that the space-time of
Solar System is described by a metric with constant-curvature
background added by a static perturbation, we show that many $f(R)$
gravities are ruled out by Solar System tests.

\end{abstract}

\end{frontmatter}

\maketitle

There had been a considerable interest in the recent cosmological
discoveries including those of high redshift type Ia supernovae
\cite{snia}, and the cosmic microwave background (CMB) \cite{cmb}
combined with the large scale structure of the Universe \cite{lss}.
In short, there is now multiple evidence indicating that more than
70\% of the critical density of the universe is in the form of
so-called dark energy; there is no understanding as to what is its
origin and nature. Given the importance and challenge of this
problem, a number of authors considered the possibility that cosmic
acceleration is not due to some kind of stuff, but rather arises
from new gravitational physics. In particular, some extensions of
general relativity were shown to predict an experimentally
consistent universe evolution without the need for dark energy.


One kind of particular proposals are so-called $f(R)$ theories of
gravity, where the gravitational Lagrangian depends on an arbitrary
analytic function $f$ of the scalar curvature $R$
\cite{Carroll:2003wy,Capozziello:2005,Vollick:2003}. It was shown
recently that $f(R)$ gravity is cosmologically viable theory because
it may contain matter dominated and radiation phase before
acceleration \cite{Cappozziello:2006}.

However, due to the excellent success of general relativity (GR) in
explaining the gravitational phenomena in Solar System, every theory
of gravity that aims at explaining the accelerated expansion of the
Universe, should reproduce GR at the Solar System scale. More
recently, Erickcek \textit{et al} \cite{Erickcek} used an elegant
method to compare the theory with Solar System tests and proved that
$1/R$ theory \cite{Carroll:2003wy} is ruled out.

In this paper, on the premise that the space-time of Solar System is
described by a metric with constant-curvature background added by a
static perturbation, we show, using the method used recently by
Erickcek \emph{et al}, that many $f(R)$ gravities are, not only
$1/R$ theory, ruled out by Solar System tests.

By varying  the gravitational action of $f(R)$ gravity,
\begin{equation}
     S=\frac{1}{2 \kappa^2} \int\, \mathrm{d}^4x\, \sqrt{-g} f(R) + \int\, \mathrm{d}^4x\, \sqrt{-g}
     {\mathcal L}_{\mathrm M},
\label{action}
\end{equation}
with respect to the metric $g_{\mu\nu}$,  we can obtain the field
equation
\begin{eqnarray}
{f'(R)}R_{\mu\nu}- \frac{1}{2}{f(R)}g_{\mu\nu}
 + \left(g_{\mu\nu}\nabla_{\alpha}\nabla^{\alpha} -
\nabla_{\mu}\nabla_{\nu}\right) f(R)=\kappa^2 T_{\mu\nu},
\label{fieldeq}
\end{eqnarray}
where primes denote derivatives with respect to $R$. Contracting
Eq.~(\ref{fieldeq}) with the inverse metric yields
\begin{equation}
     \Box f'(R)+\frac{1}{3}\triangle(R)f'(R) =  \frac{\kappa^2}{3} T ,
\label{trace}
\end{equation}
where $\triangle(R)\equiv\left[R-\frac{2f(R)}{f'(R)}\right]$ and
$T\equiv g^{\mu\nu}T_{\mu\nu}$ is the trace of energy-momentum
tensor.

For the constant-curvature vacuum solution, \emph{i.e.}, $T=0$ and
$\nabla_\mu R=0$, we have the equation for $R$
\begin{equation}\label{Rsolutoin}
f'(R)R-2f(R) =0.
\end{equation}
It is reasonable to assume that there exists a solution for
Eq.(\ref{Rsolutoin}), $R = R_0 > 0$, corresponding to the de Sitter
spacetime with Hubble parameter $H^2 = R_0/12$ and
$R_0=\sqrt{3}\mu^2$ for $f(R)=R-\mu^4/R$ theory. According to
present experiment, we have $R_0 \sim H^2 \sim 10^{-56}$ cm$^{-2}$
and $\sqrt{R_0} r \ll1$ everywhere in the Solar System. This
discussion can generalized to any $f(R)$ theory.  Note that if
Eq.(\ref{Rsolutoin}) has no such a solution, the corresponding
$f(R)$ theory should be ignored because such theory can not explain
the current acceleration of the universe without dark energy.

According to Ref.\cite{Erickcek}, the densities and velocities in
the Solar System are sufficiently small that the spacetime in the
Solar System can be treated as  a small perturbation to the de
Sitter spacetime,
\begin{eqnarray}
     ds^2 =-\left[1+a(r)- H^2 r^2\right] dt^2
     + \left[1+b(r)-H^2 r^2\right]^{-1}dr^2 + r^2 d\Omega^2,
\label{eqn:ourmetric}
\end{eqnarray}
where the metric-perturbation variables $a(r),b(r) \ll1$. Clearly,
for $a=b=0$ it describes the de Sitter spacetime with cosmological
constant $\Lambda=3H^2$.

In the Newtonian limit suitable for the Solar System, the pressure
$p$ is negligible compared to the energy density $\rho$, and
therefore $T=-\rho$. Eq.(\ref{trace}) can be reduced as
\begin{equation}
L[f']=-\frac{\kappa^2}{3}\rho,
\end{equation}
where the differential operator on the left-hand side is given by
\begin{equation}\label{operator}
L[f']\equiv\nabla^2f'(R)+\frac{1}{3}\triangle(R)f'(R),
\end{equation}
where $\nabla^2$ is the flat-space Laplacian operator.

For illustrative purposes, we take
$f(R)=R-\frac{\mu^{2+2\alpha}}{R^{\alpha}}$ and then
\begin{equation}
L[f']=\mu^2\nabla^2\left(\frac{\mu^2}{R}\right)^{\alpha}
+\frac{1}{3}R\left[(2+2\alpha)\left(\frac{\mu^2}{R}\right)^{\alpha+1}-1\right].
\end{equation}
In this case, vacuum solution $R_0=(\alpha+2)^{1/(\alpha+1)}\mu^2$
which is an ordinary point of the function.

In the complex analysis, one can define an analytic function $f$ as
follows. Let $f: A\rightarrow \mathbb{C}$ where $A\subset\mathbb{C}$
is an open set. Then $f$ is said to be differentiable in the complex
sense at $z_0\in A$ if
\begin{equation}
f'(z_0)\equiv \lim _{z\rightarrow z_0}\frac{f(z)-f(z_0)}{z-z_0}
\end{equation}
exists. $f$ is said to be analytic on $A$ if $f$ is complex
differentiable at each $z_0\in A$. However, there is an alternate
way to define an analytic function. A function $f$ is analytic iff
it is locally representable as a convergent power series. This
series is called the Taylor series of $f$. One also wants to study
the series representation of a function that is analytic on a
deleted neighborhood; that is, a function that has an isolated
singularity. The resulting series, called the Laurent series. Now,
we investigate that $f(R)$ is a function with an isolated
singularity at $R=R_s$.

Obviously, $R_0\neq R_s$ if there is non-zero vacuum solution
$R=R_0$ for Eq.(\ref{Rsolutoin}). Therefore, $R_0$ is an ordinary
point of $f(R)$ at the deleted neighborhood. We have the Taylor
series $f(R)=\sum_{n=0}^{\infty}\frac{f^{(n)}(R_0)(\delta
R)^n}{n!}$, where the perturbation $\delta R=R-R_0$.  The operator
(\ref{operator}) can be rewritten as
\begin{equation}
L[f']=\sum_{n=0}^{\infty}\left[\frac{A_n}{n!}(\delta
R)^{n-1}+\frac{B_n}{n!}(\delta R)^{n}+\frac{C_n}{n!}(\delta
R)^{n+1}\right],
\end{equation}
where
\begin{equation}
A_n=n
f^{(n+2)}(R_0)\left({\frac{\mathrm{d}R}{\mathrm{d}r}}\right)^2,
\end{equation}
\begin{equation}
B_n=f^{(n+2)}(R_0)\left(\frac{\mathrm{d}^2R}{\mathrm{d}r^2}+\frac{1}{r}\frac{\mathrm{d}R}{\mathrm{d}r}\right),
\end{equation}
and
\begin{equation}
C_n=R_0f^{(n+2)}(R_0)-2f^{(n+1)}(R_0).
\end{equation}
Here, we have already used the formula
\begin{equation}
\nabla^2(\delta R)^n=n(n-1)\frac{\mathrm{d}R}{\mathrm{d}r}(\delta
R)^{n-2}+n\left(\frac{\mathrm{d}^2R}{\mathrm{d}r^2}+\frac{2}{r}\frac{\mathrm{d}R}{\mathrm{d}r}\right)(\delta
R)^{n-1}.
\end{equation}

Clearly, expect for the case of $f^{(n)}(R_0)=0$ ($n=2,3,\cdots$),
corresponding to $f(R)=R-\Lambda$, the $C_n$ term is negligible
compared to the $A_n$ and $B_n$ terms, then $L(f')=\nabla^2f'$ in
the Newtonian limit. Note that this neglected term is a actually
Yukawa-type correction which was pointed out recently by many
authors \cite{Olmo}. Therefore, Eq.~(\ref{trace}) can be reduced to
\begin{equation}
    \nabla^2 f'(R) = - \frac{\kappa^2}{3} \rho.
\label{linearized}
\end{equation}
The solution of above equation for $r>R_\odot$ is
\begin{equation}\label{frp}
f'(R) = \frac{\kappa^2M}{12 \pi r}[1+{\mathcal O}(\sqrt{R_0}
r)]+\frac{2f(R_0)}{R_0},
\end{equation}
where $M=\int_0^{R_\odot}4\pi r^2\rho \mathrm{d}r$. Here we have
considered the de Sitter background.

Combing Eq.(\ref{fieldeq}) with Eq.(\ref{trace}) we obtain equations
for the $tt$ and $rr$ components of the Ricci tensor up to the
leading order,
\begin{eqnarray}
     R^t_t &=& 3H^2 +
     \frac{2}{f'(R)} \nabla^2 f'(R), \label{Rtt} \\
     R^r_r &=& 3H^2
     -\frac{2}{r f'(R)}\frac{\mathrm{d} f'(R)}{\mathrm{d}r}. \label{Rrr}
\end{eqnarray}
On the other hand, according to metric given by
Eq.~(\ref{eqn:ourmetric}), one have the $tt$ and $rr$ component of
the Ricci tensor(to linear order in small quantities)
\begin{eqnarray}
     R^t_t &=& 3H^2 - \frac{1}{2} \nabla^2 a(r), \label{Rtt1} \\
     R^r_r &=& 3H^2-\frac{1}{r}\frac{\mathrm{d} b(r)}{\mathrm{d}r}-\frac{1}{2}\frac{\mathrm{d}^2 a(r)}{\mathrm{d}r^2}. \label{Rrr1}
\end{eqnarray}
After comparing the corresponding equations above and using
Eq.(\ref{frp}), we obtain the solutions of $a(r)$ and $b(r)$ for
$r>R_\odot$
\begin{eqnarray}
     a(r) &=& -\frac{\kappa^2R_0}{6\pi f(R_0)}\frac{ M}{r}, \label{a} \\
     b(r) &=& -\frac{\kappa^2R_0}{12\pi f(R_0)}\frac{ M}{r} .\label{b}
\end{eqnarray}
Therefore, the effective Newton's constant is determined by
$G=\frac{\kappa^2R_0}{12\pi f(R_0)}$.

In the parameterized post-Newtonian formulism, the metric for the
Solar System can usually be rewritten as the isotropic form
\begin{eqnarray}\nonumber
ds^2=&-&\left(1-2\frac{GM}{\rho}+2\beta\frac{G^2M^2}{\rho^2}+\dots\right)\mathrm{d}t^2\\
&+&\left(1+2\gamma\frac{GM}{\rho}+\dots\right)\left(\mathrm{d}\rho^2+\rho^2\mathrm{d}\Omega^2\right),
\end{eqnarray}
where $\beta$ and $\gamma$ are dimensionless parameters. It is well
known that via the transformation $r\equiv\rho(1+\gamma
GM/\rho+\dots)$, the isotropic metric can be translated into the
Schwarzshild form \cite{Weinberg}
\begin{eqnarray}\label{24}\nonumber
ds^2=&-&\left(1-2\frac{GM}{r}+2(\beta-\gamma)\frac{G^2M^2}{r^2}+\dots\right)\mathrm{d}t^2\\
&+&\left(1+2\gamma\frac{GM}{r}+\dots\right)\mathrm{d}r^2-r^2\mathrm{d}\Omega^2.
\end{eqnarray}
It is easy to find that the PPN parameter $\gamma=1/2$ from
Eqs.(\ref{a}), (\ref{b}) and (\ref{24}). More generally, it is not
difficult to extend above discussion to the theory in which $f(R)$
has two or more isolated singularities.

General relativity predicts precisely that light deviates from
rectilinear motion near the Solar. The $f(R)$ gravity also predicts
that the photon is deflected through an angle
\begin{equation}
\triangle\varphi=\frac{4 G
M}{R_{\odot}}\frac{1+\gamma}{2}=1.75''\left(\frac{1+\gamma}{2}\right).
\end{equation}
So there is a reveal difference between the two theories.  In fact,
recent measurements in Solar System have pinned down $\gamma$ into $
1 + (2.1 \pm 2.3) \times 10^{-5}$ \cite{Bertotti:2003rm}. Hence,
$f(R)$ gravity is actually ruled out by the Solar System tests
except for theories in which $f(R)$ takes $R-\Lambda$ or some pretty
special forms. As an example of latter, $f(R)=R-
\mu^2\sin(\mu^2/(R-\Lambda))$, which has a non-isolated singularity.
It should be stressed that above discussions are on the premise that
the space-time of Solar System is described by a metric with
constant-curvature background added by a static perturbation.

\section*{Acknowledgements}
This work is supported by National Natural Science Foundation of
China under Grant No. 10473007 and No. 10503002 and Shanghai
Commission of Science and technology under Grant No. 06QA14039.

\end{document}